\documentclass[12pt]{article}
\usepackage[utf8]{inputenc}
\usepackage[paperwidth=8.5in,paperheight=11in,margin=1in,footskip=0.25in]{geometry}
\usepackage{tikz}
\usepackage{caption}
\usepackage{hyperref}
\usetikzlibrary{shapes,arrows}
\usepackage{graphicx}
\graphicspath{ {./Images/} }
\usepackage{amsmath}
\usepackage{amssymb}
\usepackage{ dsfont }
\usepackage{multirow}
\usepackage{amsthm}
\usepackage{mathtools}
\usepackage{forest}
\usepackage{setspace}
\usepackage{footnote}
\usepackage{bm}
\usepackage[tableposition=top]{caption}
\usepackage{xcolor}
\usepackage[ruled,vlined]{algorithm2e}

\usepackage{mathtools}

\DeclareMathOperator{\E}{\mathbb{E}}

\DeclareMathOperator*{\sampVar}{SampleVar}

\DeclareMathOperator{\pwaic}{\mathit{p}_{\text{WAIC}}}
\DeclareMathOperator{\W}{WAIC}
\DeclareMathOperator{\lp}{lppd}

\usepackage{natbib}

\begin{document}

\title{A numerically stable online implementation and exploration of WAIC through variations of the predictive density, using NIMBLE}

\author{Joshua E. Hug\\Department of Statistics, University of California\\Berkeley, California, USA \\ \\
  Christopher J. Paciorek\\Department of Statistics, University of California\\Berkeley, California, USA\\paciorek@stat.berkeley.edu}

\maketitle

\newpage

\begin{abstract}
We go through the process of crafting a robust and numerically stable online algorithm for the computation of the Watanabe-Akaike information criteria (WAIC). We implement this algorithm 
in the NIMBLE software. The implementation 
is performed in an online manner and does not require the storage in memory of the complete samples from the posterior distribution. This algorithm allows
the user to specify a specific form of the predictive density to be used in the computation of WAIC, in order to cater to specific prediction goals. 
We then comment and explore via simulations the use of different forms of the predictive density in the context of different predictive goals. We find that when using marginalized predictive densities, 
WAIC is sensitive to the grouping of the observations into a joint density.
\end{abstract}

\section{Introduction}

After building several candidate Bayesian models, we often seek to evaluate the models based on predictive accuracy. A common 
method to assess this is the Watanabe-Akaike information criterion (WAIC) \citep{watanabe}. \citet{gelman} discuss WAIC and compare to other forms of predictive information 
criteria. They conclude that WAIC is a fast and computationally convenient alternative to true leave-one-out cross-validation. \citet{vehtari} compare WAIC and Pareto-smoothed importance sampling leave-one-out cross validation. WAIC is typically calculated fairly simply 
using a pointwise log predictive density and is obtained from samples of the posterior distribution of interest. The predictive density is not 
unique, and different prediction goals require the use of different forms of the predictive density. The current implementation of WAIC in the NIMBLE software \citet{nimble}
is limited in two ways. The first limitation is that, apart from changing the model definition, there is no way for the user to utilize a specific predictive density form 
when they are interested in specific predictive goals. \citet{ariyo} , specifically discuss 
the differences between marginal and conditional predictive density use with linear mixed models for longitudinal data. They conclude that the marginal criterion 
is often more effective at choosing the correct model in this context. The second limitation is that all computation of WAIC occurs after Markov chain Monte Carlo posterior (MCMC) sampling and requires the storage of all 
MCMC samples in memory to compute. We solve both problems by describing an online algorithm that computes WAIC as MCMC sampling occurs, not requiring the storage in 
memory of any MCMC samples. This algorithm also allows the user to specify the form of the predictive density they are interested in so that proper contextual model selection may be performed. Finally, we perform simulations using this algorithm to assess the performance of the different forms of WAIC.

\section{Estimating out of sample pointwise predictive accuracy using WAIC}

Suppose that we are given data partition elements, $y_1,\dots, y_M$, that are conditionally independent when conditioned on some set of unknown parameters, $\theta$.
We can express our likelihood $p(y|\theta)=\prod_{m=1}^M p(y_m|\theta)$. Further suppose we impose a prior distribution, $p(\theta)$. This yields a posterior distribution,
$p(\theta|y)$, and a posterior predictive distribution, $p(\tilde y_m|y)= \int p(\tilde y_m|\theta)p(\theta|y)d\theta$. Here $\tilde y_m$ represents 
future or alternative data, and the collection $\{y_m, \ m=1,\ldots, M\}$, represents an entire new or alternative dataset. We label the true 
data generating process as $f$.

We define the expected log pointwise predictive density (elpd) for a new dataset: 

\begin{align*}
    \text{elpd}&=\sum_{m=1}^M \E_f(\log(p(\tilde y_m|y))\\
    &=\sum_{m=1}^M \int \log(p(\tilde y_m|y)) f(\tilde y_m) d\tilde y_m.
\end{align*}
This represents one possible measure of a model's out of sample predictive performance.

In machine learning this is sometimes known as the mean log predictive density. We can note that $\text{elpd}=\sum_{m=1}^M\left(\E_f(\log(f(\tilde y_m))) - D_{KL}\left(f(\tilde y_m)||p(\tilde y_m|y)\right)\right)$, where $D_{KL}$ is the Kullback–Leibler divergence. The true data generating process $f$ is unknown. Since the new or alternative data  are random we average over them.\\

In practice, we only have access to a single dataset, and we seek to estimate elpd with this dataset.  We approximate the elpd using the true log 
pointwise predictive density (lppd),

\begin{align*}
    \lp&=\sum_{m=1}^M \log(p(y_m|y))\\
    &=\sum_{m=1}^M \log \left(\int p(y_m|\theta)p(\theta|y)d\theta\right).
\end{align*}
We obtain this approximation by removing the integral with respect to $f$ (the expectation), and evaluating the posterior predictive density at our current dataset. We are treating our dataset as a single draw from the true data generating process $f$, so we estimate this expectation with the log predictive density evaluated at our available dataset.\\

Since the integral over the parameters is typically intractable, we compute the lppd 
 using samples from the posterior distribution as

\begin{align}
    \widehat\lp = \sum_{m=1}^M \log \left(\frac{1}{S} \sum_{s=1}^S p(y_m|\theta^s)\right).
\end{align}

In this case we are evaluating the likelihood at each of the $s = 1, \dots, S$ , Markov Chain Monte Carlo samples from 
the posterior distribution: $p(\theta|y)$.\\

Since the lppd is computed using existing data it is a biased estimate for new data. We correct this bias by subtracting 
an estimate of the effective number of parameters, denoted as the $\pwaic$, 

\begin{align}
    \pwaic &= \sum_{m=1}^M \sampVar_s(\log(p(y_m|\theta^s))),
\end{align}
where we define $\sampVar_s(x_s) = \frac{1}{S-1} \sum_{s=1}^S \left(x_s - \frac{1}{S}\sum_{s=1}^S x_s\right)^2$\\

There is a mean based formulation of $\pwaic$, discussed in \citet{gelman} and denoted $\pwaic_1$, but we do not discuss this formulation as it is not as commonly used as (2).

Our WAIC-driven estimate of the elpd is 

\begin{align*}
    \widehat{\text{elpd}}_{\text{WAIC}} &= \widehat\lp - \pwaic.
\end{align*}

For historical reasons, measures of prediction accuracy such as AIC, DIC and WAIC are typically defined based on the deviance scale which is the 
log predictive density multiplied by $-2$. We can then define WAIC as

\begin{align*}
    \W &= -2(\widehat\lp - \pwaic).
\end{align*}

For notational convenience we write $\widehat\lp$ as $\lp$, and only compute (1) in practice.

A simple form of WAIC is implemented in many popular software packages for Bayesian inference (NIMBLE, Stan, PyMC3) \citep{nimblepack,stan,pymc}.\\

Methods such as AIC \citep{akaike} and DIC \citep{spiegelhalter} exist as alternatives to WAIC
to approximate the elpd. However, these methods only utilize a point estimate of the unknown parameters. For a more fully Bayesian approach we
use the entire posterior distribution to evaluate the predictive performance. Certain asymptotic properties of AIC only hold for regular statistical models. WAIC on the other hand extends useful asymptotic properties to singular (non-regular) statistical models, which generally include hierarchical models \citep[see][for mathematical rigour]{watanabe}.

\section{Choice of posterior predictive density}

The equations for the lppd (1) and $\pwaic$ (2) contain the same loosely-defined building blocks. We have not clearly defined how to obtain each of 
the terms used in their computation, and we now illustrate the flexibility in user choice when computing the elements of the WAIC. The first user choice is the 
form of the predictive density $p(y_m|\theta^s)$. For instance,
consider the case where $\theta = \{\theta_1,\theta_2\}$. We may be interested in the predictive distribution that solely depends on 
$\theta_1$.\\

 The second case for possible user choice is in the data partition, 
$y_1,\dots, y_M$, as it is not obvious as to what each $y_m$ can or should be represented as. Options include, for instance, a single data point, 
or a vector of data points each belonging to some hierarchical level or group. Other forms of correlation such as temporal or spatial make the 
question of the data partition even more unclear. We explore both options and potential motivations for their specific use 
in the sections that follow.

\subsection{Marginalized log predictive density}

Consider the log predictive density, henceforth called the conditional log predictive density,
 evaluated at some data partition element, $y_m$:

\begin{align}
    \text{conditional log predictive density} &= \log p(y_m|\theta).
\end{align}
This term appears in both the lppd and the $\pwaic$. If $\theta$ is multidimensional then we can rewrite (3) as:

\begin{align*}
    \text{conditional log predictive density} &= \log p(y_m|\theta_1,\theta_2 ),
\end{align*}
where we have that $\theta = \{\theta_1,\theta_2\}$. For a particular predictive goal we may not wish to condition on every element in the parameter vector. In this case we look to compute a marginalized predictive density,
which involves conditioning on only a subset of the full parameter vector $\theta$:

\begin{align}
    \text{marginal log predictive density} &= \log p(y_m|\theta_1)  \\
    &= \log \int p(y_m,\theta_2|\theta_1) d\theta_2\tag*{}\\
    &=\log \int p(y_m|\theta_1,\theta_2) p(\theta_2|\theta_1)d \theta_2\tag*{}.
\end{align}

Consider the following simple hierarchical model as motivation for the use of the marginal predictive density.

\begin{align*}
    \phi &\thicksim p(\eta)\\
    \mu_j| \phi &\thicksim p(\phi),  \ j = 1, \dots , J \tag{A}\\
    z_{j,i} | \mu_j &\thicksim N(\mu_j, 1),  \ j = 1, \dots , J, \  i = 1, \dots, n_j
\end{align*}

In this scenario we have that $\theta = \{\mu_1, \mu_2, \dots, \mu_J, \phi \}$. If we were to use the conditional log predictive density as in (3), 
then our predictive density would be conditioned on all of the group means: $ \mu_1, \dots, 
\mu_J$. This is useful if our goal is to predict data from existing
groups $\{\mu_j\}$, as the density of a data point conditioned on an existing group mean makes sense. However, if instead we seek 
to test the predictive ability of the model on data from new groups then this conditioning does not make sense. It is reasonable to 
assume that certain models would be more effective at predicting data from existing group means, and vice versa. If our goal is to predict new data 
from new groups then it would make sense to marginalize out the group means, and in accordance with our previous notation we would have that 
$\theta_1 = \phi$, and $\theta_2 = \{\mu_1,\dots,\mu_J\}$.\\

Anytime we are interested in computing WAIC with some marginalized predictive density, we call this the \textit{marginal WAIC}. If we do not marginalize, 
we call this the \textit{conditional WAIC}.

When we consider marginalization it is very important 
to also consider the data partition used to compute the WAIC. We describe in section 10 how certain specifications of the data partition when using marginal 
predictive densities can lead to WAIC choosing an incorrect model.

\subsection{Data Partitions}

Consider a dataset with data points: $z_1, z_2, \dots , z_n$. We now consider choosing a partition (or grouping) of length $M \leq n$ which we denote as 
$y_1, \dots, y_M$, where each $y_m$ contains one or more of the $z_i$ elements. 
Often, we are simply interested in how well our model predicts new individual data points, which are typically scalars. In this sense 
the data partition elements are the individual, typically scalar data points. Take the simple model 

\begin{align*}
    \mu &\thicksim p(\phi)\\
    z_j | \mu & \overset{\text{ind}}\thicksim N(\mu, 1), \ j = 1, \dots, n.
\end{align*}

The most natural data partition here is that we take $M=n$ and we have that 
$y_m=z_i$ for $m=i$.

Suppose we observe the slightly more complicated hierarchical model described in A.

Now the choice isn't as obvious as before since we could consider again $y_m=z_{j,i}$ for $m=1,...,\sum_{i=1}^J n_j$=, but this choice doesn't reflect the natural structure of this model.
 We could instead think of 
this model as being pointwise in the sense that each of the $j = 1, \dots J$ groups whose observations have identical mean $\mu_j$ are 
the "points" of interest. I.e., we could take $M = J$ and have $y_m = \{z_{j,i}: j=m, j= 1, \dots , n_j \}$. In this case $y_m$ would be a vector of all the data points 
with same mean $\mu_j$, and $p(y_m|\theta^s)$ would be a joint density over the vector $y_m$.\\

In general, our first option is to partition with what will be known as an \textit{ungrouped} or \textit{individual} pointwise selection:
each data point as defined within the model is taken as the partition.  Otherwise 
we may have a custom \textit{grouped} partition selection such as that of the grouped data in the hierarchical model given above. 
We explore later the meaning behind this choice and why a user may wish to choose a specific data partition structure.\\

In general, we can take the inputted data as it is defined via the model likelihood as

\[ z_1, \dots, z_n.\]

The default data partition can be written as

\[y_1=\{z_1\}, y_2=\{z_2\}, \dots, y_M=\{z_n\} ,\] 
which we will call the \textit{ungrouped WAIC}. If we use any grouping of the 
data to form different partitions, then we call this a \textit{grouped WAIC}. For instance a generic partition of the 
data may be

\[ y_1=\{z_1, \dots, z_{n_1}\}, y_2=\{z_{n_1+1}, \dots , z_{n_2}\}, \dots, y_M=\{z_{n_{K-1}+1}, \dots, z_{n_K}\}. \]

With conditionally independent data, logical data partitions often present themselves naturally due to the structure of the model. In cases 
with correlated data like time series or spatial data, it is unclear how to properly partition the data. In these cases using an ungrouped pointwise selection (summing 
over the individual observations) is not a clear answer since we still have not addressed the underlying correlation structure. The question of partitioning 
with time series data is further addressed in Section 11.

\section{Introduction to NIMBLE}

NIMBLE is a system for programming statistical algorithms for general
model structures within R \citep{nimblepack,R}. In particular, for our purposes NIMBLE extends the BUGS \citep{gilks} language and allows the 
creation of model objects for tasks such as calculating log probability values and sampling from a posterior distribution via MCMC. 
NIMBLE currently provides an option to calculate WAIC after sampling from a posterior distribution for a user-
defined model. The WAIC currently implemented is the ungrouped conditional WAIC.
 
Most internal computations in NIMBLE occur over nodes, which are essentially vertices in a model's graphical representation. The data nodes 
are, as their name implies, nodes for which we have data. Therefore these nodes are fixed by a user input and cannot be 
simulated over. NIMBLE distinguishes two main kinds of nodes, stochastic and deterministic nodes. Stochastic nodes are those 
defined by a distributional assignment, for instance, 

\[ y \thicksim \text{dnorm}(\text{mean},\text{sd = standard deviation}).\]
Deterministic nodes are determined through deterministic calculations but can be dependent on stochastic elements, such as 

\[ c \gets y+3 .\]

In our case we will only be concerned with two particular types of nodes within a 
NIMBLE model. These are the latent parameter nodes and the data nodes. The latent parameter nodes are those parameters chosen to marginalized out when using marginal WAIC.  Data nodes are fixed and can never be simulated over; we can only fetch 
values of log probabilities at these nodes. The latent nodes are not fixed, and we are able to simulate random draws of the parameters at these nodes 
using the \textit{simulate} function call. For our purposes the \textit{calculate} call determines the log probability values of whichever data
nodes are inputted into the function call. We can then fetch the log probability at a single data node or a sum of log probabilities of various data nodes together  via the \textit{getLogProb} call.
 Therefore we can \textit{simulate} over certain stochastic nodes and then \textit{calculate} and \textit{getLogProb}  to obtain the log probability values. This sequence 
of calls is the main method to obtain the necessary log predictive density values needed to calculate the WAIC via NIMBLE.

\section{The need for an online algorithm} 
The current NIMBLE implementation of the WAIC involves first sampling via MCMC from the desired posterior distribution and then computing the WAIC 
using these samples after they are stored in memory. More precisely, the posterior samples for all stochastic parent nodes of all the data nodes 
need to be saved to perform this computation. In addition, the current NIMBLE implementation only offers ungrouped conditional WAIC.
Computation of a conditional WAIC requires the computation and storage 
of a matrix containing the log predictive density values, whose dimension is $M$ $\times$ $S$, where $M$ is the number of data 
partition elements and $S$ is the number of posterior MCMC samples. Both the storage of the posterior samples as 
well as the creation of this log probability matrix can be memory intensive. Rather than computing the WAIC after the sampling from the posterior, we 
propose an algorithm to compute the WAIC in an online manner during the sampling. This algorithm works by storing and updating only the necessary quantities computed from from the MCMC samples in order to compute (1) and (2). We update each of these elements as we sample, immediately after an iteration in which the entire vector of 
parameters has been sampled. This ensures that for the purpose of WAIC calculation, we do not need to store any MCMC samples in memory beyond the sample at the current iteration. The following 
section describes the algorithm for this computation.

\section{Bit by Bit: Constructing an online algorithm}
In order to simplify 
notation and elucidate more complex aspects of the algorithm, we rewrite equations (1) and (2) as:

\begin{align}
    \text{WAIC}&= -2\left(\text{lppd}-p_{\text{WAIC}} \right), \tag*{}\\
    \tag*{}\\
    \text{lppd} &= \sum_{m=1}^M\log\left(\frac{1}{S}\sum_{s=1}^S \exp(h_m(\theta^s))\right), \\
    \tag*{}\\
    p_{\text{WAIC}} &= \sum_{m=1}^M \sampVar_s(h_m(\theta^s)),
\end{align}

where we define $h_m(\theta^s)=\log(p(y_m|\theta^s))$.\\
At first we will assume that 
the $h_m(\theta^s)$ values are given to us easily from NIMBLE, which is only true when computing the ungrouped conditional WAIC. 
For marginal 
predictive densities, computing the proper form of $h_m(\theta^s)$ becomes its own challenge in an online manner. We cover the procedure of computing the correct 
user-specified predictive density after describing
the computation of each element of the WAIC as if the $h_m(\theta^s)$ are given.

\subsection{The outer sum}
In both (5) and (6) the outer sum over the data partition elements, $\sum_{m=1}^M$, cannot be computed until we have finished sampling from 
the posterior. All calculations that follow are computed at each element of a data partition vector, $m$.The algorithms below 
only describe the computation at a single element of the data partition since these are vectorized computations. To finalize the computation after the 
MCMC sampling has completed, we sum over all the partition elements, $\sum_{m=1}^M$, to compute the finalized (1) and (2). 

\subsection{The $\pwaic$}

The $\pwaic$ presents the first challenge in computing the WAIC in an online fashion. We must compute the sample variance over the $h_m(\theta^s)$ evaluated at the $s=1,\dots,S$  MCMC samples. 
Algorithms for online variance computation have been studied extensively \citep[e.g.,][]{ling}, and we choose to 
employ Welford's online variance algorithm \citep{welford}.

Algorithm 1 shows how to calculate the sample variance of a set of values, $x_1, x_2, \dots , x_S$, without storing the full set of values.

\begin{algorithm}[H]
    \SetAlgoLined
    count $\gets$ 0 \\
    M2 $\gets$ 0 \\
    $\overline x$ $\gets$ 0\\
    \For{s in 1:S}{
        count $+=$ 1\\
        $\delta_1$ $\gets$ $x_s - \overline x$\\
        $\overline x$ $+=$ $\delta_1/$count\\
        $\delta_2$ $\gets$  $x_s - \overline x$\\
        M2 $+=$ $\delta_1 \cdot \delta_2$
    }
    Finalization Step: $\sampVar_s$ $\gets$ M2$/($count$-1)$
    \caption{Welford's online variance}
\end{algorithm}

Using algorithm 1 we can compute $\sampVar_s(h_m(\theta^s))$ in an online fashion. This algorithm is applied to every element of the data 
partition in a vectorized manner. This requires the storage of two vectors of length $M$.

\subsection{The lppd and the online logSumExp}

The equation for computation of the lppd for a single data partition element is

\begin{align}
    \log\left(\frac{1}{S}\sum_{s=1}^S \exp(h_m(\theta^s))\right)= -\log(S)+\log \left(\sum_{s=1}^S \exp(h_m(\theta^s))\right),
\end{align}
where $-\log(S)$ is a constant so we only describe the calculation of the second term.

Because this computation involves taking the log of a sum of exponentiated terms, we do not wish to compute this term directly due to risk of numerical underflow. This issue is typically solved by
implementing a logSumExp trick. The existing NIMBLE implementation solves this 
issue via this logSumExp calculation:

\begin{align*}
    \log \left(\sum_{s=1}^S (\exp(h_m(\theta^s))\right)&= \max_s(h_m(\theta^s)) + \log \left(\sum_{s=1}^S (\exp(h_m(\theta^s)-\max_s(h_m(\theta^s)))\right)
\end{align*}
This generic 
logSumExp trick requires us to have access to the maximum of the elements over which we are summing, $s=1,\dots,S$. However, in an online implementation, we do not store these values and only have access to one at a time. To solve this we propose a 
method to perform the logSumExp trick in an online fashion.

We will store the current maximum value of the log probability 
and update the summation to attempt to contain any possible future underflow. Consider iterating through $h_m(\theta^s), s=1,\dots, S$ as we compute the summation. The first 
value seen, $h_m(\theta^1)$, is treated as the current maximum value, since we have observed no other values. We begin the summation denoting $h_m(\theta^1)$ (the current maximum) as $v_1$.
Suppose that $h_m(\theta^{n_1})$ is the first value that we observe greater than $h_m(\theta^1)$. Denote the partial sum inside the log in (7) up to this point as $\Omega_1$:

\begin{align*}
    \Omega_{1} = \exp(h_m(\theta^1)-v_1)+ \exp(h_m(\theta^2)-v_1) + \dots + \exp(h_m(\theta^{n_1 -1})-v_1)
\end{align*}

Now that we have discovered a new maximum value, $h_m(\theta_{n_1})$, we define $v_2 =h_m(\theta_{n_1})-v_1$. We note that $v_1+v_2$ is now the maximum value observed.
 We define the second summation as follows by subtracting a new maximum from the exponent, until we observe the next new maximum at $n_2$.

\begin{align*}
    \Omega_{2} =  \exp(h_m(\theta^{n_1})-(v_1 + v_2))+ \exp(h_m(\theta^{n_1+1})-(v_1 + v_2)) + \dots \\
    + \exp(h_m(\theta^{n_2-1})-(v_1 + v_2))
\end{align*}

We repeat the process through all the $s=1, \dots, S$ iterations until we have all of our partial sums, $\{\Omega_1, \dots, \Omega_c\}$, as well as the partial maximums,
$\{v_1, \dots, v_c\}$.

We can always analytically recover the full sum in (7):

\begin{align*}
   \log\left( \exp\left(\sum_{i=1}^c v_i\right)IS\right)&=\sum_{i=1}^c v_i + \log(IS)\\
   IS &=\sum_{i=1}^c \left(\exp\left(-\sum_{j=i+1}^c v_j\right)\Omega_i\right)
\end{align*}

We only show the term $\exp\left(\sum_{i=1}^c v_i\right)$ in order to explain the algorithm. In practice we never evaluate this term and only evaluate the right side of the equation.

Equivalently we can show this expansion, illustrated here assuming only three maximum are ever found:

\begin{align*}
    &\log \left(\sum_{s=1}^S \exp(h_m(\theta^s))\right)  \\
    &= v_1 + \log\left ( \sum_{s=1}^S \exp(h_m(\theta^s)-v_1)) \right )\\
    &=v_1 + v_2+\log \left( \left [e^{-v_2}\sum_{s=1}^{n_1} \exp(h_m(\theta^s)-v_1) + \sum_{n_1+1}^S \exp(h_m(\theta^s)-(v_1+v_2))  \right] \right) \\
    &= v_1+v_2+v_3+ \log \left (\left [e^{-v_3}\left(e^{-v_2}\sum_{s=1}^{n_1} \exp(\exp(h_m(\theta^s)-v_1) \right. \right. \right.\\
   &\left. \left. \left. + \sum_{n_1+1}^S \exp(h_m(\theta^s)-(v_1+v_2))  \right) + \sum_{s=n_2+1}^S \exp(h_m(\theta^s)-(v_1+v_2+v_3))\right] \right)
\end{align*}

In the actual online algorithm, we only need to store the current maximum, $\sum_{i=1}^{c'}v_i$, denoted CurrentMax, as well as the current scaled value of the inner sum IS, denoted CurrentSum.
Every time a new maximum is found, we scale the inner sum by the new partial maximum $v_i$ and then update the CurrentMax, only storing the inner sum and the currentMax.\\
\begin{algorithm}[H]
    \SetAlgoLined
    \For{s in 1:S}{
        \If{s = 1}{
            CurrentMax $\gets$ $h_m(\theta^s)$\\
            CurrentSum $\gets$ 1
        }\ElseIf{$h_m(\theta^s)> \text{CurrentMax}$}{
            NewV $\gets$ $h_m(\theta^s)-$ CurrentMax\\
            CurrentMax $+=$ NewV\\
            CurrentSum $\gets$ CurrentSum $\cdot$ $\exp(-\text{NewV})+1$
        }\Else{
            CurrentSum $\gets$ CurrentSum + $\exp(h_m(\theta^s)-\text{CurrentMax})$
        }
    }
    Finalization Step: Final $\gets$ CurrentMax + $\log\left(\text{CurrentSum}\right) - \log(S)$ 
    \caption{Online LogSumExp trick}
\end{algorithm}

Using Algorithm 2, we compute (7) in an online fashion. This algorithm is applied to every element of the data 
partition in a vectorized manner. This requires the storage of two vectors of length $M$.

\section{Computing the user-specified predictive density values}

The above process describes being given the predictive density $h_m(\theta^s)=\log(p(y_i|\theta^s))$ values. These values are only simple to access 
in the case of ungrouped conditional WAIC. For either grouped partitions or for marginal WAIC, we must compute a non-standard 
predictive density value to be used in the above algorithms. 

\subsection{Computing the marginal predictive density values}

We will first explore computing the marginal predictive density. Suppose we are able 
to divide our parameters as $\theta = \{\theta_1,\theta_2\}$ and we are interested in marginalizing out the parameters $\theta_2$. Then we can expand (4) further as:

\begin{align}
    \text{marginal log predictive density} &= \log p(y_m|\theta_1)\tag*{}\\
    &=\log \int p(y_m|\theta_1,\theta_2) p(\theta_2|\theta_1)d \theta_2\tag*{}\\
    &\approx \log \left(\frac{1}{K}\sum_{k=1}^K p\left(y_m|\theta_1,\theta_2^{k}\right)\right),\ \ \theta_2^{k} \thicksim p(\theta_2|\theta_1^s)\tag*{}\\
    &= \log \left(\frac{1}{K}\sum_{k=1}^K \exp\left(h_m(\theta_1,\theta_2^k)\right)\right) 
\end{align}

Thus, in order to approximate the marginal density, we sample the latent parameters conditioned on posterior draws of the other parameters
$\theta_2^{(k)} \thicksim p(\theta_2|\theta_1^s), k=1,\ldots,K$, and we compute the marginal density via Monte Carlo simulation.\\

All density values in NIMBLE are on the log scale so we have to exponentiate before summing. As with the previous algorithms 
we must compute (8) at each of the $M$ data partition elements. The full equations for the marginal WAIC elements are:

\begin{align}
    \lp = \sum_{m=1}^M \log \left(\frac{1}{S} \sum_{s=1}^S \left(\frac{1}{K}\sum_{k=1}^K \exp\left(h_m(\theta_1^s,\theta_2^k)\right) \right)\right)
\end{align}

\begin{align}
    \pwaic &= \sum_{m=1}^M \sampVar_s\left(\log \left(\frac{1}{K}\sum_{k=1}^K \exp\left(h_m(\theta_1^s,\theta_2^k)\right)\right) \right)
\end{align}

Given these formulas, there are multiple potential avenues to perform the computation of (8) at every data partition element, $m=1,\dots, M$. One possible approach 
involves looping over the data partition elements and computing the marginal density (8) at each data partition element. This method would require the simulation 
of $\theta_2$, $K$ times, at each data partition element, for every MCMC sample. This requires a total of $M\cdot K \cdot S$ simulations of $\theta_2$. This is computationally inefficient
since we should not need to resimulate the $K$ values of $\theta_2$ at every data partition element. Our chosen procedure does not simulate at each data partition element and requires only 
$K\cdot S$ simulations.  The 
issue with this approach is that we are computing a logSumExp in (8) without access to all $K$ entries in the summation. We only have access to the 
one simulation of the latent parameter $\theta_2^k$ at a time and need to update the entire data partition vector. We have already solved this problem with Algorithm 2 for online 
logSumExp. Using this method preserves the vectorization over the data partition.

Our chosen procedure is:

\begin{enumerate}
    \item Sample a value of the latent parameter $\theta_2$.
    \item Update the logSumExp for (8) using Algorithm 2 at every data partition element $m=1,\dots, M$.
    \item Repeat steps 1 and 2 $K$ times.
    \item Finalize the logSumExp to obtain (8) at $m=1,\dots, M$ and use these values for the computation of equations (5) and (6).
    \item Repeat the previous three steps at every MCMC iteration $s=1,\dots,S$.
\end{enumerate}

\subsection{Computing grouped predictive density values}

Second, when working with a user-provided grouped WAIC, computing the predictive density over a vector-valued partition $y_m$ is not difficult since NIMBLE will automatically 
output the sum of log probabilities when provided with a vector of data nodes. Our implementation works with this by allowing users 
to provide specific partitions using the node names of the data inputs, with which we correctly specify the partition in the software for the computation 
of the predictive density values. The algorithms given previously do not change; all prior equations treat $y_m$ as a vector, and $p(y_m|\theta)$ as a joint density.

\section{Putting it together: The full online algorithm}

\subsection{Full algorithm}

The full online algorithm is a combination of the previous parts. A general overview of the algorithm is:

\begin{enumerate}
    \item Receive the $s$th posterior sample during MCMC.
    \item Compute the user-specified  log predictive density values at each data partition element (Sections 7.1-7.2).
    \item Update necessary stored quantities for lppd (Section 6.3).
    \item Update necessary stored quantities for $\pwaic$ (Section 6.2).
    \item Upon completion of MCMC sampling, aggregate terms for lppd and $\pwaic$ and return the WAIC.
\end{enumerate}

In the following algorithm we denote the conditional parameters of interest as $\theta_1$ and the latent nodes as $\theta_2$. We use $K$ as in equation (8) as the number of Monte Carlo itertations used to approximate the integral for the marginal density. If we are using conditional WAIC then $\theta_2$ does not have any elements ($\theta = \theta_1$), and $K=1$. 

\begin{algorithm}[H]
    \SetAlgoLined
    \If{marginal WAIC}{
    \For{k in 1:K}{
            \begin{enumerate}
                \item[]
                \item[] Simulate the latent nodes $\theta_2^k$.
            \end{enumerate}
        
        \For{m in 1:M}{
                \begin{enumerate}
                    \item[] Update the online logSumExp to compute predictive density\\
                    as in Section 7.1 
                    using Algorithm 2 (omit the finalization step\\
                    and outer loop).
                \end{enumerate}

        }
    } 
    Finalize Algorithm 2 for the online logSumExp and obtain $h_m(\theta^s), \ m = 1,\dots, M$.\\
    }
    \If{conditional WAIC}{
        Obtain $h_m(\theta^s), \ m = 1,\dots, M$ log conditional density directly through NIMBLE.
    }
     \For{m in 1:M}{
        \begin{enumerate}
            \item Update lppd element vectors using Algorithm 2 (omit the finalization step\\
            and outer loop).
            \item Update $\pwaic$ element vectors using Algorithm 1 (omit the finalization step\\
            and outer loop).
        \end{enumerate}
    }
    \caption{Online WAIC update given a single MCMC sample}
\end{algorithm}

This algorithm is executed at every iteration of the MCMC sampler, so we omit the loops over the MCMC samples as they are written in Algorithms 1 and 2.
 Following the completion of the sampling we can finalize to compute the lppd, $\pwaic$ and the WAIC. Both 
of these are simply vectorized computations, but we illustrate this as a for loop over the data partition. The notation $a^m$ denotes the $m$th entry in the 
vector. \\ \\
\begin{algorithm}[H]
    \SetAlgoLined
    \For{m in 1:M}{
        \begin{enumerate}
            \item Finalize $\pwaic$ sample variance computation:\\
            $\sampVar_s^m$ $\gets$ $\text{M2}^m$ $/($count$-1)$ .
            \item Finalize lppd computation:\\
            $\text{lppd}^m$ $\gets$ $\text{CurrentMax}^m $+ $\log\left(\text{CurrentSum}^m\right) - \log(S)$.
        \end{enumerate}
    }
    lppd $\gets$ $\sum_{m=1}^M \text{lppd}^m $\\
    \ \\
    $\pwaic$ $\gets$ $\sum_{m=1}^M \sampVar_s^m $\\
    \ \\
    WAIC $\gets$ $-2(\lp - \pwaic)$
    \caption{Finalize WAIC after MCMC sampling}
\end{algorithm}

\subsection{Variance of mWAIC values}

When we compute the marginal predictive density, we introduce variance into the WAIC computation through the Monte Carlo integral approximation in (8).
Ideally we would like to compute the variance of this approximation. This would require us to be able to compute the variance of the equations (9) and (10) with respect to the randomness 
introduced from the $\theta_2$. Further this would require the variance of a log sum for (9) and a sample variance for (10). This is analytically challenging.
 In order to facilitate some analysis by the user, NIMBLE automatically computes 
the WAIC, $\pwaic$ ,and lppd at $0.25K, \ 0.5K, \ 0.75K$ where K is by default set to $1000$ but can be changed by the user. Therefore the user 
can observe if any of the terms seem to be unstable due to low K. This also requires NIMBLE to store four times as many vectors as stated in the earlier sections when computing the marginal WAIC. Interestingly, our empirical testing indicates that much of the variability occurs in the $\pwaic$ computation rather than the $\lp$. We do not have an explanation for this phenomenon and it is potentially worth exploring in the future.

\section{Simulations}

The goal of our simulations is to investigate using different forms of WAIC on simple models in hopes of better understanding how to choose what form to use in practice. Particularly 
we experiment with using marginal WAIC on both grouped and ungrouped data, in light of the Ariyo et al. (2019) results. Our first example is a hierarchical model
with a clear grouping structure. Our second example is a stochastic volatility time series model with correlation that 
does not have a clear grouping structure. 

\subsection{Simple random intercept hierarchical model}
We initially explore the different forms of WAIC with a simulation of a simple random intercept hierarchical model. The 
true data generating process is given by:

\begin{align*}
    b_j| \mu, \tau \overset{\text{ind}}\thicksim& N(\mu, \tau), \ j = 1, \dots, J\\
    y_{j,i}| b_j, \sigma \overset{\text{ind}}\thicksim& N(b_j,\sigma), \ j = 1, \dots, J, \ i=1,\dots,n_j
\end{align*}

We fit three models to compare; the first, model H, is the true model:

\begin{align*}
    \mu &\thicksim p(\phi_1)\\
    \sigma &\thicksim p(\phi_2)\\
    \tau &\thicksim p(\phi_3)\tag{H}\\
    b_j| \mu, \tau \overset{\text{ind}}\thicksim& N(\mu, \tau), \ j = 1, \dots, J\\
    y_{j,i}| b_j, \sigma \overset{\text{ind}}\thicksim& N(b_j,\sigma), \ j = 1, \dots, J, \ i=1,\dots,n_j
\end{align*}

The second model forces the $\tau$ parameter to be very small, reducing the true variation in the group means $b_j$. We denote this
incorrect model F:

\begin{align*}
    \mu &\thicksim p(\phi_1)\\
    \sigma &\thicksim p(\phi_2)\tag{F}\\
    b_j| \mu \overset{\text{ind}}\thicksim& N(\mu,0.01), \ j = 1, \dots, J\\
    y_{j,i}| b_j, \sigma \overset{\text{ind}}\thicksim& N(b_j,\sigma), \ j = 1, \dots, J, \ i=1,\dots,n_j
\end{align*}

Our third model ignores any hierarchical structure and is denoted S:

\begin{align*}
    \mu &\thicksim p(\phi_1)\\
    \sigma &\thicksim p(\phi_2)\tag{S}\\
    y_{j,i}| \mu, \sigma \overset{\text{ind}}\thicksim& N(\mu,\sigma), \ j = 1, \dots, J, \ i=1,\dots,n_j
\end{align*}

Model S has no latent variables so we have that marginal and conditional WAIC are identical. In addition these marginal and conditional values should 
be nearly identical to those of model F. Model F forces the $\tau$ to be $0.01$, essentially removing all variation in the $b_j$ values. Model S removes all the variation 
and simply assumes a single mean across all observations.

\subsection{Results}

In our simulations we use $\tau =0.5$, $\sigma = 1$ and $\mu = 2$. We compute the WAIC with 500 different simulated datasets for each of the two simulations. We provide the means of the WAIC, lppd, and $\pwaic$ averaged over the simulated datasets, for each variation of predictive density. We also provide the Monte Carlo standard error to demonstrate the variation of the mean over the datasets. We consider 
four types of WAIC for this model. In accordance with our earlier description, for both the H and F model we compute ungrouped conditional  WAIC, grouped conditional WAIC, ungrouped marginal WAIC, and grouped marginal WAIC. For the grouped 
WAIC we define each data partition element as $C_m = \{y_{m,i}:i=1,\ldots,n_m\}$. Therefore each partition element is all the data with the same group mean. 
For computing the marginal predictive density we define our latent variables ($\theta_2$ in earlier notation) as $\{b_1, \dots, b_J\}$, or the set of 
group means. If our predictive goal is to predict new observations from new groups, then we should use this marginal predictive density since 
it is not conditioned on a new data point belonging to an existing group. For marginal WAIC we used $K = 1000$ for the Monte Carlo integration 
approximation. We use 5000 samples following the initial 500 being used for burn-in. We graphically checked for proper mixing. \\

\subsection{Hierarchical Simulation 1}

Our first set of 500 simulations uses $J=20$, $n_j = 100$ for all $j$. Table \ref{hiermeans1} reflects our theoretical knowledge that model S and model F should be very similar models. We also 
see that for model F, the conditional and marginal WAIC are similar due to the forced limited variation in the group means. Table \ref{hiersel1} indicates that the ungrouped marginal WAIC never selects the correct model. We explore the 
reason for this in Section 10. Besides the ungrouped marginal all other forms of WAIC select the correct model. Table \ref{hiermcer1} shows that there is some variability but the results we find are robust to Monte Carlo simulation error.

\begin{table}
    \centering
    \caption{Hierarchical simulation 1 means over 500 simulated datasets.}\label{hiermeans1}
    \begin{tabular}{|lr|ccc|}
      \hline
     WAIC type & model type  & mean(WAIC) & mean(lppd) & mean($\pwaic$) \\ 
      \hline
      \multirow{3}{10em}{grouped  conditional} &  H & 5690.99 & -2834.50 & 10.99 \\ 
       & F & 6126.09 & -3036.44 & 26.61 \\ 
       & S & 6129.39 & -3040.63 & 24.07  \\
       \hline
       \multirow{3}{10em}{ungrouped  conditional} & H  & 5696.51 & -2828.04 & 20.21 \\ 
        & F & 6099.71 & -3047.71 & 2.14  \\ 
        & S & 6105.69 & -3050.86 & 1.99 \\ 
        \hline
        \multirow{3}{10em}{grouped marginal}& H & 5745.81 & -2869.60 & 3.30 \\ 
        & F& 6126.20 & -3039.22 & 23.88 \\ 
        & S & 6129.39 & -3040.63 & 24.07  \\
        \hline
        \multirow{3}{10em}{ungrouped marginal}& H & 6178.08 & -3055.07 & 33.97 \\ 
         & F & 6105.72 & -3050.85 & 2.01 \\ 
         & S & 6105.69 & -3050.86 & 1.99 \\ 
         \hline
    \end{tabular}
    \captionsetup{singlelinecheck=false}
    \caption*{Model S has identical marginal and conditional values.}
    \end{table}

\begin{table}
    \centering
    \caption{Hierarchical simulation 1 Monte Carlo error over 500 simulated datasets.}\label{hiermcer1}
    \begin{tabular}{|lr|ccc|}
      \hline
     WAIC type & model type  & mean(WAIC) & mean(lppd) & mean($\pwaic$) \\ 
      \hline
      \multirow{3}{10em}{grouped  conditional} & H & 2.84 & 1.42 & 0.01 \\ 
          &F  & 6.49 & 2.96 & 0.34 \\ 
      & S & 6.52 & 2.99 & 0.32 \\ 
       \hline
       \multirow{3}{10em}{ungrouped  conditional} &  H & 2.84 & 1.42 & 0.01  \\ 
          &  F& 6.17 & 3.09 & 0.003 \\ 
        & S& 6.22 & 3.11 & 0.003\\ 
        \hline
        \multirow{3}{10em}{grouped marginal}& H & 2.85 & 1.42 & 0.07 \\ 
            & F &  6.50 & 2.97 & 0.32\\ 
         & S & 6.52 & 2.99 & 0.32 \\ 
        \hline
        \multirow{3}{10em}{ungrouped marginal}& H& 9.02 & 3.28 & 2.01  \\ 
           & F& 6.22 & 3.11 & 0.003\\ 
         & S& 6.22 & 3.11 & 0.003\\ 
         \hline
    \end{tabular}
    \captionsetup{singlelinecheck=false}
    \caption*{Model S has identical marginal and conditional values.}
    \end{table}

        \begin{table}
            \centering
             \caption{Hierarchical simulation 1 proportion of correct models selected.}\label{hiersel1}
            \begin{tabular}{|c|c|}
              \hline
              WAIC type & proportion of correct model selected  \\ 
              \hline
              ungrouped conditional& 1 \\
              ungrouped marginal & 0 \\
             grouped conditional& 1\\
             grouped marginal& 1\\
            \hline
            \end{tabular}
            \end{table}

We cannot assess Monte Carlo error of the proportion of correct models selected since there is no variability.

\subsection{Hierarchical Simulation 2}

Our second set of 500 simulations uses $J=40$, $n_j = 60$ for all $j$. The substantive results are equivalent to those of the first hierarchical simulation.

\begin{table}
    \centering
    \caption{Hierarchical simulation 2 means over 500 simulated datasets.}\label{hiermeans2}
    \begin{tabular}{|lr|ccc|}
      \hline
     WAIC type & model type  & mean(WAIC) & mean(lppd) & mean($\pwaic$) \\ 
      \hline
      \multirow{3}{10em}{grouped  conditional} &  H & 6859.08 & -3398.70 & 30.84 \\ 
       & F & 7352.48 & -3663.62 & 12.62 \\ 
       & S & 7353.96 & -3665.92 & 11.07 \\
       \hline
       \multirow{3}{10em}{ungrouped  conditional} & H  & 6873.15 & -3381.49 & 55.09 \\ 
        & F & 7341.70 & -3668.67 & 2.18  \\ 
        & S & 7344.63 & -3670.32 & 1.99 \\ 
        \hline
        \multirow{3}{10em}{grouped marginal}& H & 6964.85 & -3478.91 & 3.51 \\ 
        & F& 7352.47 & -3665.20 & 11.03 \\ 
        & S& 7353.96 & -3665.92 & 11.07   \\
        \hline
        \multirow{3}{10em}{ungrouped marginal}& H & 7365.56 & -3670.59 & 12.20 \\ 
         & F & 7344.64 & -3670.32 & 2.00 \\ 
         & S & 7344.63 & -3670.32 & 1.99 \\ 
         \hline
    \end{tabular}
    \captionsetup{singlelinecheck=false}
    \caption*{Model S has identical marginal and conditional values.}
    \end{table}

\begin{table}
    \centering
    \caption{Hierarchical simulation 2 Monte Carlo error over 500 simulated datasets.}\label{hiermcer2}
    \begin{tabular}{|lr|ccc|}
      \hline
     WAIC type & model type  & mean(WAIC) & mean(lppd) & mean($\pwaic$) \\ 
      \hline
      \multirow{3}{10em}{grouped  conditional} & H & 3.21 & 1.61 & 0.01\\ 
          &F  & 4.74 & 2.31 & 0.08 \\ 
      & S & 4.74 & 2.32 & 0.08  \\ 
       \hline
       \multirow{3}{10em}{ungrouped  conditional} &  H & 3.21 & 1.61 & 0.05 \\ 
          &  F& 4.68 & 2.34 & 0.003 \\ 
        & S& 4.69 & 2.34 & 0.003\\ 
        \hline
        \multirow{3}{10em}{grouped marginal}& H & 3.15 & 1.58 & 0.06 \\ 
            & F& 4.74 & 2.32 & 0.08\\ 
         & S & 4.74 & 2.32 & 0.08 \\ 
        \hline
        \multirow{3}{10em}{ungrouped marginal}& H& 4.81 & 2.35 & 0.08  \\ 
           & F& 4.69 & 2.34 & 0.003 \\ 
         & S& 4.69 & 2.34 & 0.003\\ 
         \hline
    \end{tabular}
    \captionsetup{singlelinecheck=false}
    \caption*{Model S has identical marginal and conditional values.}
    \end{table}

        \begin{table}
            \centering
            \caption{Hierarchical simulation 2 proportion of correct models selected.}\label{hiersel2}
            \begin{tabular}{|c|c|}
              \hline
              WAIC type & proportion of correct model selected  \\ 
              \hline
              ungrouped conditional& 1 \\
              ungrouped marginal & 0 \\
             grouped conditional& 1\\
             grouped marginal& 1\\
            \hline
            \end{tabular}
            \end{table}

\section{Marginal WAIC sensitivity to partitioning}

We observe in our results that the ungrouped marginal WAIC does not select the correct model. Consider a simplified description of model H; this model 
has a random effect $b_j$ for every group.

\begin{align*}
    b_j| \mu, \tau \thicksim& N(\mu, \tau), \ j = 1, \dots, J \tag{H}\\
    y_{j,i}| b_j, \sigma\thicksim& N(b_j,\sigma), \ j = 1, \dots, J, \ i=1,\dots,n
\end{align*}

Now consider a model in which every data point has a separate random effect:

\begin{align*}
    b_{j,i}| \mu, \tau \thicksim& N(\mu, \tau), \ j = 1, \dots, J \ , i=1,\dots,n \tag{$S_1$} \\
    y_{j,i}| b_{j,i}, \sigma \thicksim& N(b_{j,i},\sigma), \ j = 1, \dots, J, \ i=1,\dots,n
\end{align*}

We can rewrite $S_1$ into an equivalent model as:

\begin{align*}
    y_{j,i}| \mu , \sigma, \tau \thicksim& N\left(\mu,\sqrt{\sigma^2 +\tau^2}\right), \ j = 1, \dots, J, \ i=1,\dots,n \tag{$S_2$}
\end{align*}

Model $S_2$, and therefore model $S_1$, is also equivalent to model S. Since there are no latent parameters with model $S_2$ but there are with model $S_1$ this implies
that in this case the ungrouped marginal and the ungrouped conditional for models $S_1$ and $S_2$ are identical. This is irrespective of the true value of $\tau$.

We can also note that if we were to use the ungrouped marginal for H or for $S_1$ we should obtain the same results since our marginal predictive density at a single data partition element
has the form:

\begin{align*}
  \text{(H) ungrouped marginal predictive density}  &= \log \left(\frac{1}{K}\sum_{k=1}^K p(y_{j,i}|b_j^k))\right), \ b_j^k \thicksim N(\mu,\tau) \\
  \text{($S_1$) ungrouped marginal predictive density}  &= \log \left(\frac{1}{K}\sum_{k=1}^K p(y_{j,i}|b_{j,i}^k))\right), \ b_{j,i}^k \thicksim N(\mu,\tau)
\end{align*}

These are the same because when we use ungrouped WAIC, we completely ignore the grouping structure. We generate 
the same predictive densities even when we are comparing two fundamentally different models H and $S_1$. Thus, use of ungrouped marginal WAIC to compare models H and $S_2$ is equivalent to comparing models $S_1$ and $S_2$. 
Given that $S_1$ and $S_2$ are equivalent, we cannot expect ungrouped marginal WAIC to be able to choose the correct model H. Put otherwise, the distinction between models H and either $S_1$ or $S_2$ is the dependence amongst observations in a group and not the individual predictions. Ungrouped marginal WAIC does not account for dependence between observations.

However if we partition with $\bm{y_m} = \{y_{j,i}:j=m\}, \ m= 1, \dots, M$ (i.e., partition the data into groups which share the same group mean),
then we have that the marginal predictive densities have the forms:

\begin{align*}
    \text{(H) grouped marginal predictive density}  &= \log \left(\frac{1}{K}\sum_{k=1}^K p(\bm{y_m}|b_j^k))\right), \ b_j^k \thicksim N(\mu,\tau), \\
    \text{($S_1$) grouped marginal predictive density}  &= \log \left(\frac{1}{K}\sum_{k=1}^K p(\bm{y_m}|b_{j,1}^k, \dots, b_{j,n_j}))\right),\\
    \ &b_{j,i}^k \thicksim N(\mu,\tau).
  \end{align*}

These are no longer identical, since the second generates a new group mean, $b_{j,i}$, for each element of the $\bm{y_m}$ vector. This indicates that the marginal density is sensitive to the partition. If we wish to obtain marginal WAIC results that make sense, we have to respect the natural 
partition induced by the model hierarchy.\\

 Since WAIC calculation relies on the data and cannot make use of the unknown latent variables,
 a comparison of models with different latent structure must in some way make use of the aspects of the data (such as use of joint density values or conditioning on latent group variables) that could reflect latent structure if present.
 Ungrouped marginal WAIC does not do this.\\
 
  It is interesting to note that \citet{ariyo} implicitly use grouping in their calculation of marginal WAIC for linear mixed models for longitudinal data. Because they have Gaussian data, they are able to compute a closed form marginal multivariate normal density. These multivariate densities group all measurements of a single subject, as we did for our hierarchical model example. While in our hierarchical example it would also be possible to compute a closed form for the marginal density, we note that NIMBLE does not have the capability to recognize this automatically. Our implementation of marginalization approximates an integral via simulation and applies to general models that do not have a closed form solution.\\

 In light of these results about grouping with marginal WAIC, one might consider how to approach correlation without discrete groups such as time series data.

\section{Time Series Data}

\subsection{Stochastic volatility data}

Stochastic volatility models model the variance of an asset $y_t$ as following a latent stochastic process in time. The true data generating process is given by:

\begin{align*}
    h_1|\phi,\mu,\sigma &\thicksim N\left(\mu, \frac{\sigma}{\sqrt{1-\phi^2}}\right),\\
    h_t|h_{t-1},\phi,\mu,\sigma &\thicksim N\left(\mu+ \phi(h_{t-1}-\mu),\sigma\right),\\
    y_t|h_t &\thicksim N\left(0, \exp\left(\frac{h_t}{2}\right)\right).
\end{align*}

We observe the $y_t$ and attempt to model the variance through a latent process $h_t$. The $h_t$ are modeled 
as an AR(1) process. The true model is denoted P.

\begin{align*}
    \sigma &\thicksim p(\eta_1)\\
    \mu &\thicksim p(\eta_2) \\
    \phi &\thicksim p(\eta_3)\\
    h_1|\phi,\mu,\sigma  &\thicksim N\left(\mu, \frac{\sigma}{\sqrt{1-\phi^2}}\right)\tag{P}\\
    h_t|h_{t-1},\phi,\mu,\sigma  &\thicksim N\left(\mu+ \phi(h_{t-1}-\mu),\sigma\right), \ t=2,\dots, T\\
    y_t|h_t &\thicksim  N\left(0, \exp\left(\frac{h_t}{2}\right)\right), \ t=1,\dots, T
\end{align*}

Our second model gets rid of the AR(1) relationship between the $h_t$'s by forcing $\phi = 0 $, and we denote this model Z.

\begin{align*}
    \sigma &\thicksim p(\eta_1)\\
    \mu &\thicksim p(\eta_2) \\
    h_1|\mu,\sigma  &\thicksim N\left(\mu, \sigma\right)\tag{Z}\\
    h_t|\mu,\sigma  &\thicksim N\left(\mu,\sigma\right), \ t=2,\dots, T\\
    y_t|h_t &\thicksim  N\left(0, \exp\left(\frac{h_t}{2}\right)\right), \ t=1,\dots, T
\end{align*}

Our third model removes all latent structure, and we model the observations as independent
realizations from a normal distribution with mean zero. This model is denoted I.

\begin{align*}
    \sigma \thicksim p(\eta_1)\\
    y_t|\sigma \thicksim N(0,\sigma) \tag{I}
\end{align*}

\subsection{Results}

We simulate 300 stochastic volatility datasets. We use $\phi = 0.95$, $\sigma = 0.25$, $T=200$, and $\mu = -1.02$. These parameter values are similar to those used in the Stan user manual \citep{stan}. We use $K=3000$ iterations for marginal WAIC. Since grouping is not obvious in this case, we somewhat arbitrarily choose several different grouping options for the models presented. We use 5000 samples following the initial 500 being used for burn-in. We graphically checked for proper mixing. Again, there is some variability in the WAIC values, but the standard error estimates indicate that the results are robust to Monte Carlo simulation error.

\begin{table}
    \centering
     \caption{Stochastic volatility simulation means over 300 simulated datasets.}\label{svmeans}
    \begin{tabular}{|lc|ccc|}
      \hline
     WAIC type & model type  & mean(WAIC) & mean(lppd) & mean($\pwaic$) \\ 
      \hline
      \multirow{3}{10em}{ungrouped  conditional} &  P & 382.55 & -172.83 & 18.45  \\ 
       & Z & 398.26 & -173.87 & 25.26\\ 
       & I & 421.52 & -208.97 & 1.79\\
       \hline
       \multirow{3}{10em}{grouped (2 data points) conditional} & P  & 382.75 & -172.71 & 18.67 \\ 
        & Z  & 398.63 & -173.41 & 25.90 \\ 
        & I & 421.82 & -208.85 & 2.06 \\ 
        \hline
        \multirow{3}{10em}{grouped (10 data points) conditional}& P& 383.88 & -171.95 & 19.99\\ 
        & Z & 398.70 & -171.61 & 27.74\\ 
        & I & 423.63 & -208.20 & 3.62 \\ 
        \hline
        \multirow{3}{10em}{grouped (20 data points) conditional}& P & 383.51 & -171.01 & 20.75\\ 
         & Z & 406.10 & -170.00 & 33.05 \\ 
         & I & 425.16 & -207.76 & 4.83 \\ 
         \hline
         \multirow{3}{10em}{grouped (all 200 data points) conditional}& P & 398.18 & -166.06 & 33.03 \\ 
         & Z & 453.45 & -162.45 & 64.27 \\ 
         & I & 420.38 & -209.68 & 0.51 \\ 
         \hline
        \multirow{3}{10em}{ungrouped marginal} &  P & 426.80 & -205.23 & 8.17 \\ 
       & Z & 411.65 & -203.61 & 2.22 \\
       & I & 421.52 & -208.97 & 1.79\\
       \hline
       \multirow{3}{10em}{grouped (2 data points) marginal} & P  & 420.62 & -203.38 & 6.93 \\ 
        & Z  & 412.41 & -203.65 & 2.56 \\ 
        & I & 421.82 & -208.85 & 2.06 \\ 
        \hline
        \multirow{3}{10em}{grouped (10 data points) marginal}& P& 408.77 & -198.70 & 5.68\\ 
        & Z & 412.61 & -202.88 & 3.42 \\ 
        & I & 423.63 & -208.20 & 3.62 \\ 
        \hline
        \multirow{3}{10em}{grouped (20 data points) marginal}& P & 400.62 & -196.48 & 3.83\\ 
         & Z & 417.15 & -202.57 & 6.00\\ 
          & I & 425.16 & -207.76 & 4.83 \\ 
         \hline
         \multirow{3}{10em}{grouped (all 200 data points) marginal}& P & 504.82 & -192.80 & 59.61 \\ 
         & Z & 617.19 & -205.45 & 103.15 \\ 
         & I & 420.38 & -209.68 & 0.51 \\ 
       \hline
    \end{tabular}
    \captionsetup{singlelinecheck=false}
    \caption*{Model I has identical marginal and conditional values.}
    \end{table}

\begin{table}
    \centering
    \caption{Stochastic volatility Monte Carlo error over 300 simulated datasets.}\label{svMCer}
    \begin{tabular}{|lc|ccc|}
      \hline
     WAIC type & model type  & se(WAIC) & se(lppd) & se($\pwaic$) \\ 
      \hline
      \multirow{3}{10em}{ungrouped  conditional} &  P & 4.16 & 2.12 & 0.30  \\ 
       & Z & 4.41 & 2.19 & 0.54\\ 
       & I & 4.67 & 2.32 & 0.04\\
       \hline
       \multirow{3}{10em}{grouped (2 data points) conditional} & P  & 4.18 & 2.12 & 0.29 \\ 
        & Z  & 4.57 & 2.23 & 0.64 \\ 
        & I & 4.67 & 2.32 & 0.06 \\ 
        \hline
        \multirow{3}{10em}{grouped (10 data points) conditional}& P& 4.23 & 2.14 & 0.45\\ 
        & Z & 4.38 & 2.17 & 0.58 \\ 
        & I & 4.71 & 2.31 & 0.13  \\ 
        \hline
        \multirow{3}{10em}{grouped (20 data points) conditional}& P & 4.16 & 2.13 & 0.36\\ 
         & Z & 6.06 & 2.19 & 2.04  \\ 
         & I & 4.75 & 2.30 & 0.19 \\ 
         \hline
         \multirow{3}{10em}{grouped (all 200 data points) conditional}& P & 4.29 & 2.14 & 0.92 \\ 
         & Z & 5.71 & 2.20 & 2.04  \\ 
         & I & 4.65 & 2.33 & 0.003 \\ 
         \hline
        \multirow{3}{10em}{ungrouped marginal} &  P & 9.41 & 2.51 & 3.87\\ 
       & Z & 14.82 & 2.26 & 6.51 \\
       & I & 4.67 & 2.32 & 0.04\\
       \hline
       \multirow{3}{10em}{grouped (2 data points) marginal} & P  & 4.69 & 2.23 & 0.30  \\ 
        & Z  & 4.49 & 2.22 & 0.29 \\ 
        & I & 4.67 & 2.32 & 0.06 \\ 
        \hline
        \multirow{3}{10em}{grouped (10 data points) marginal}& P& 5.40 & 2.16 & 1.36 \\ 
        & Z & 4.51 & 2.22 & 0.11  \\ 
        & I & 4.71 & 2.31 & 0.13  \\ 
        \hline
        \multirow{3}{10em}{grouped (20 data points) marginal}& P & 4.29 & 2.14 & 0.10\\ 
         & Z & 4.68 & 2.22 & 0.36\\ 
         & I & 4.75 & 2.30 & 0.19 \\ 
         \hline
         \multirow{3}{10em}{grouped (all 200 data points) marginal}& P & 9.41 & 2.51 & 3.87 \\ 
         & Z & 14.82 & 2.26 & 6.51 \\ 
         & I & 4.65 & 2.33 & 0.003 \\ 
       \hline
    \end{tabular}
    \captionsetup{singlelinecheck=false}
    \caption*{Model I has identical marginal and conditional values.}
    \end{table}

        \begin{table}
            \centering
            \caption{Stochastic volatility proportion \& standard error of models correctly selected.}\label{svsel}
            \begin{tabular}{|c|c|c|}
              \hline
             WAIC type & proportion & se(proportion)\\
             \hline
              ungrouped conditional & 0.917 & 0.016 \\
              grouped (2 data points) conditional & 0.9 & 0.017 \\
             grouped (10 data points) conditional & 0.89& 0.018\\
             grouped (20 data points) conditional & 0.887 & 0.018\\
             grouped (all 200 data points) conditional & 0.71 & 0.026\\
             \hline 
              ungrouped marginal& 0.04 & 0.011\\
              grouped (2 data points) marginal & 0.123 & 0.019 \\
             grouped (10 data points) marginal & 0.77 & 0.024\\
             grouped (20 data points) marginal & 0.873& 0.019\\
             grouped (all 200 data points) marginal & 0.043 & 0.012\\
            \hline
            \end{tabular}
            \end{table}

    Interestingly, marginal WAIC rarely chooses the correct model when using a grouping of relatively few data points. Consistent with the hierarchical example we see that the ungrouped marginal rarely selects the correct model. The proportion of the simulations in which marginal WAIC chooses the correct model increases as we increase the grouping to a moderate size. However as we move to the extreme of grouping with the entire dataset, the proportion of correct models selected decreases.  These results point to directions for  further research in regards to grouping and the marginal WAIC. It is unclear whether grouping with the entire data set makes sense in the context of WAIC. A blog post (\url{https://statmodeling.stat.columbia.edu/2014/09/25/waic-time-series/}) written in part by Aki Vehtari indicates that the motivation makes sense in principle, but WAIC relies on an asymptotic argument with respect to the summation over the data partition elements, so having $M=1$ may not give accurate results.

\section{Discussion}

We have developed an online algorithm for the computation of WAIC that allows for users of NIMBLE to select the form of predictive density used in the calculation. In addition, we show that different forms of the predictive density lead to potentially different conclusions when using WAIC. Particularly, we demonstrate that when using marginal WAIC, it is essential to use some form of grouping to obtain correct results. Our results show that this also holds for general correlation structures with unclear grouping. We have implemented this algorithm into the NIMBLE software.

\bibliographystyle{apalike}
\bibliography{refs}

\end{document}